\documentclass[preprint,aps,12pt,showpacs,nofootinbib,tightenlines]{revtex4}
\usepackage{amsmath}
\usepackage{amssymb}
\usepackage{epsfig}
\usepackage{graphicx}
\def\q_slash{\not{\hbox{\kern-2.1pt $q$}}}
\def\p_slash{\not{\hbox{\kern-2.1pt $p$}}}
\def\k_slash{\not{\hbox{\kern-2.1pt $k$}}}
\def\bmaT{\left(\begin{array}{ccc}}
\def\emaT{\end{array}\right)}
\let\jnfont=\rm
\def\NPB{\jnfont Nucl.\ Phys.\ B}
\def\PLB{\jnfont Phys.\ Lett.\ B}
\def\EPJC{\jnfont Euro.\ Phys.\ J.\ C}
\def\PRD{\jnfont Phys.\ Rev.\ D}
\def\PRL {\jnfont Phys.\ Rev.\ Lett.}

\def\JHEP{\jnfont J. High \ Ener.\  Phys.}
\def\RMP{\jnfont  Rev. Mod. Phys.}

\def\CPC{\jnfont Chin. Phys. C}
\def\PLB{\jnfont Phys. Lett. B}

\newcommand{\ba}{\begin{array}}
\newcommand{\ea}{\end{array}}
\newcommand{\bd}{\begin{displaymath}}
\newcommand{\ed}{\end{displaymath}}
\newcommand{\beq}{\begin{equation}}
\newcommand{\eeq}{\end{equation}}
\newcommand{\bea}{\begin{eqnarray}}
\newcommand{\eea}{\end{eqnarray}}

\newcommand{\MNS}{{\text{MNS}}}

\begin{document}


\title{Probing Lepton Flavor Violation Signal via
 $\gamma\gamma\to  \bar \ell_i \ell_j$ in the
Left-Right Twin Higgs Model at the ILC}

\author{Guo-Li Liu\footnote{guoliliu@zzu.edu.cn}, Fei Wang, Kuan Xie, Xiao-Fei Guo }
\affiliation{  Department of Physics, Zhengzhou University, Henan,
450001, China}

\begin{abstract}
To explain the small neutrino masses, heavy Majorana neutrinos are introduced in the left-right twin Higgs model.
The heavy neutrinos, together with the charged scalars and the heavy gauge
bosons, may contribute large mixings between the neutrinos and the charged leptons,
which may induce some distinct lepton flavor violating processes. We will check
the $\bar \ell_i \ell_j$ ($i,~j= e,~\mu,~\tau,~i\neq j$) productions
in the $\gamma$$\gamma$ collision in the left-right twin Higgs model,
and find that the production rates may be large in some specific parameter space,
in the optimal cases even possible to be detected with reasonable kinematical cuts.
we have also shown that these collisions can constrain effectively the model parameters
such as the Higgs vacuum expectation value and the right-handed neutrino mass, etc.,
and may serve as a sensitive probe of this new physics model.
\end{abstract}

\pacs{12.60-i, 12.60. Fr, 13.66 -a}

\maketitle

\newpage
\section{ Introduction}
One of the problems of the Standard Model (SM) is that
the neutrino oscillation experiments indicate that neutrinos are massive and mix with each other,
which manifestly require new physics beyond the SM \cite{oscillneutrinos} since in SM
the neutrino masses and thus Lepton Flavor Violating (LFV) couplings are missing.
The LFV signals, however,
are predicted in many new physics models, such as supersymmetry \cite{susy},
topcolor assisted technicolor models \cite{tc2-review}, little Higgs \cite{lh-review},
 Higgs triple models \cite{htm-review}, and the Left-Right Twin Higgs (LRTH) \cite{lrth-review} models, etc.

In the LRTH model, to provide the mass origin of the leptons and to explain the small neutrino masses,
right-handed heavy neutrinos are introduced. These right-handed heavy neutrinos can realize the mixings of the neutrinos with the leptons,
which can induce LFV processes at the proposed International Linear Collider (ILC)\cite{ilc-project},
such as the decay $\mu\to e \gamma$ \cite{0711.1238}.
We will in this paper will discuss the $\bar \ell_i \ell_j$ ($i,~j= e,~\mu,~\tau,~i\neq j$) productions via the $\gamma\gamma$
collision in the LRTH model.

Due to its rather clean environment, the ILC can be an ideal collider to probe new physics.
At the ILC, in addition to $e^+e^-$ collision, one can also realize the $\gamma\gamma$ collision \cite{rr-project} with the photon
beams generated by the backward compton scattering of incident electron- and laser-beams.

The $\gamma\gamma$ collision, however, has two advantages over the $e^+ e^-$ collision of the ILC in probing
the LFV interaction \cite{rr-advantage,0311166}.  One is that the process
$e^+ e^- \to \bar \ell_i \ell_j$ occurs only via s-channel, and the rates are
suppressed by the photon propagator and the neutral gauge boson
propagator. On the contrary, the process $\gamma \gamma \to \bar \ell_i \ell_j$  is
free of this. Another is that the backgrounds of the $e^+ e^-$ collision
may be not so easy to suppress\cite{rr-advantage}. Since the $\gamma \gamma$ collision
may be free of many SM irreducible backgrounds, the LFV productions in $\gamma\gamma$ collision
are suitable for detecting the new physics models.

We in this work will study the LFV processes $\gamma\gamma\to \bar \ell_i \ell_j$
($i\neq j$ and $\ell_{i} = e,~\mu,~\tau$) induced by the gauge bosons $W^\pm$,  $W_{H}^\pm$ and
charged scrlars $\phi^\pm$ in LRTH models, at the same time, the heavy neutrinos entering the loop.
we will find that, due to the existence of the heavy neutrinos, the production in the LRTH model
have different properties and rich phenomenology.

This paper is organized as follows. In Sec. II we briefly review the lepton sector of LRTH model
and give the couplings involved in our calculation. In Sec. III, we will discuss the contributions
from the gauge bosons and the charged scalars.
In Sec.IV, on the base of the former discussion, we will show the parameter constraints
 related to the processes.
Finally, conclusions are given in Sec. V.

\section{the lepton sector of the LRTH model and the relevant couplings}
In the LRTH model \cite{lrth-review,0711.1238,0611015-su}, with the global symmetry $U(4) \times U(4)$,
the Higgs field and the twin Higgs in the fundamental representation of each $U(4)$
 can be written as $H$ = $( H_{L}, H_{R} )$ and $\hat{H}$ = $( \hat{H}_{L},\hat{H}_{R} )$, respectively.
After each Higgs develops a vacuum expectation value (VEV),
\begin{equation}
<H> = (0,0,0,f),\quad <\hat{H}> = (0,0,0,\hat{f}),
 \label{vevs}
\end{equation}
the global symmetry $U(4)\times U(4)$ breaks to $U(3)\times U(3)$, with the gauge group
$ SU(2)_L \times SU(2)_R \times U(1)_{B-L}$ down to the SM $U(1)_Y$.
After the breaking, there are six massive gauge bosons left:
the SM $Z$ and $W^\pm$, and extra heavier bosons, $Z_{H}$ and $W_H^\pm$. And
eight scalars are left: one neutral pseudoscalar, $\phi^0$, a pair of charged
scalars $\phi^\pm$, the SM physical Higgs $h$, and an ${\rm SU}(2)_L$ twin Higgs
doublet $\hat{h}=(\hat{h}_1^+, \hat{h}_2^0)$.

Neutrino oscillations \cite{oscillneutrinos} imply that neutrinos are massive,
and the LRTH models try to explain the origin of the neutrino masses and mass hierarchy.
Three families doublets ${\rm SU}(2)_{L,R}$ are introduced in the LRTH models to provide lepton masses,
\begin{eqnarray}
    L_{L\alpha}=-i\left(\begin{array}{c}~\nu_{L\alpha}
\\l_{L\alpha}\end{array}\right), 
    ~~~L_{R\alpha}=\left(\begin{array}{c}~\nu_{R\alpha}
\\l_{R\alpha}\end{array}\right), 
\end{eqnarray}
where the family index $\alpha$ runs from 1 to 3.

In the same way as the first two generations of quarks, the charged leptons also obtain their masses
via non-renormalisable dimension $5$ operators, which for the lepton sector can be written as
\begin{equation}
{y_l^{ij}\over \Lambda} (\bar{L}_{Li} H_L)(H_R^{\dagger}{L}_{Rj})+{y_{\nu}^{ij}\over \Lambda}
(\bar{L}_{L,i}\tau_2 H_L^*)(H_R^T\tau_2{L}_{Rj}) + {\rm {H.c.}} ,
\label{eq:Yukawalep}
\end{equation}
which will give rise to lepton Dirac mass terms $y_{\nu,l}^{ij} f^2/\Lambda$, once $H_L$ and $H_R$ acquire  VEVs.

The Majorana nature of the left- and right-handed neutrinos, however, makes one to induce
Majorana terms ( only the mass section) in dimension 5 operators,
\begin{equation}
{c_{L}\over \Lambda}\, \left( \overline{L}_{L\alpha} \tau_2 H_{L}^\dagger \right)^2+ {\rm H. c},
\qquad {c_{R} \over \Lambda}\, \left( \overline{L}_{R\alpha} \tau_2 H_{R}^\dagger \right)^2+ {\rm H. c. }~.
\label{MajoranaLR}\
\end{equation}
Once $H_L~(H_R)$ obtains a VEV, both neutrino chiralities obtain Majorana masses via
these operators, the smallness of the light neutrino masses,
however, can not be well explained.

Then, if we assume that the twin Higgs $\hat{H}_R$ (which is
forbidden to couple to the quarks to prevent the heavy top
quark from acquiring a large mass of order $y \hat f$) couples to
the right-handed neutrinos, one finds that \cite{0711.1238}
\begin{equation}
{c_{\hat{H}} \over \Lambda}\, \left( \overline{L}_{R\alpha} \tau_2 \hat{H}_{R}^\dagger \right)^2+ {\rm H. c. }~,
\label{dimfiv}
\end{equation}
which will give a contribution to the Majorana mass of the heavy right-handed neutrino,
in addition to those of  Eq.(\ref{MajoranaLR}).

 After the electroweak symmetry breaking, $H_R$ and $\hat{H}_R$ get VEVs, $f$ and $\hat{f}$ (Eq.(\ref{vevs})), respectively,
 we can derive the following seesaw mass matrix for the LRTH model in the basis ($\nu_L$,$\nu_R$):
\begin{equation}
{\cal M} = \left(\begin{array}{cc}
c {v^ 2 \over 2\Lambda}  & y_{\nu} {v f \over \sqrt{2}\Lambda} \\
y_{\nu}^{T} {v f \over \sqrt{2}\Lambda} & c {f^ 2 \over \Lambda}+ c_{\hat{H}} {\hat{f}^ 2 \over \Lambda}
\end{array}\right)~.
\label{calM}
\end{equation}
In the one-generation case there is two massive states, a heavy ($\sim \nu_R$) and a light one.
For the case that $v <  f < \hat{f}$, the masses of the two eigenstates are about
$m_{\nu_{heavy}} \sim c_{\hat{H}} {\hat{f}^ 2 \over \Lambda}$ and $m_{\nu_{light}} = {c v^ 2 \over 2\Lambda} $ \cite{0711.1238}.

The Lagrangian in Eq.(\ref{eq:Yukawalep}) induces neutrino masses and the mixings of different generation leptons, which
may be a source of lepton flavour violating  \cite{0711.1238}.

We consider the contributions of the heavy gauge boson, $W_{H}$,
and the charged scalars, $\phi^\pm$, too.
The relevant vertex interactions for these processes are explicated in the followings:
\begin{equation}
\phi^-\bar l \nu_{L,R}:\frac{i}{f}(m_{l_L,\nu_R} P_L-m_{\nu_{L},l_R}P_R) V_H\sim ic_H\frac{{\hat f}^2}{\Lambda f} P_L,\label{hlv}
\end{equation}
\begin{equation}
W^-_{L,R}\bar l \nu_{L,R}:\frac{e}{\sqrt{2} s_w}\gamma_{\mu}   P_{L,R} V_H.
\label{wlv}
\end{equation}
where $V_H$ is the mixing matrix of the heavy neutrino and the leptons mediated by the charged scalars and
the heavy gauge bosons. The vertexes of $\phi^-\bar l \nu_{L,R}$ can also be expressed in the coupling constants. The
$\phi^-\bar l \nu_{R}$, for example, is also written as  $ic_H\frac{\hat f^2}{\Lambda f}P_L$
if we neglect the charged lepton masses and take $m_{\nu_h}=c_H\hat f^2/ \Lambda$.

\section{Calculations}

\subsection{The Distribution Functions in the $\gamma\gamma$ collision}
For the $\gamma \gamma$ collision at the ILC, the photon beams are
generated by the backward Compton scattering of incident electron-
and laser-beams just before the interaction point.  The events number
is obtained by convoluting the cross section with the photon beam
luminosity distribution and for the $\gamma \gamma$ collider the events
number is obtained by
\begin{eqnarray}
N_{\gamma \gamma \to   \bar \ell_i \ell_j}&=&\int  {\rm d}
\sqrt{s_{\gamma \gamma}} \frac{{\rm d}{\cal L}_{\gamma \gamma}}
{{\rm d} \sqrt{s_{\gamma \gamma}}} \hat{\sigma}_{\gamma \gamma \to
\bar \ell_i \ell_j} (s_{\gamma \gamma})\equiv {\cal L}_{e^+e^-}
~\sigma_{\gamma \gamma \to   \bar \ell_i \ell_j}(s_{e^+e^-}),
\label{definition}
\end{eqnarray}
where ${\rm d}{\cal L}_{\gamma \gamma}/{\rm d} \sqrt{s_{\gamma
\gamma}}$ is the photon beam luminosity distribution and
$\sigma_{\gamma \gamma \to   \bar \ell_i \ell_j}(s_{e^+e^-}) $, with
$s_{e^+e^-}$ being the energy-square of $e^+e^-$ collision, is
defined as the effective cross section of $ \gamma \gamma \to
\bar \ell_i \ell_j$. In optimum case, $\sigma_{\gamma \gamma \to
\bar \ell_i \ell_j} $ can be written as \cite{distribution}
\begin{eqnarray}
\sigma_{\gamma \gamma \to
\bar \ell_i \ell_j}(s_{e^+e^-})&=&\int_{\sqrt{a}}^{x_{max}} 2 z{\rm d}
z
 ~\hat{\sigma}_{\gamma \gamma \to   \bar \ell_i \ell_j} (s_{\gamma \gamma}=z^2 s_{e^+e^-})
\int_{z^2/x_{max}}^{x_{max}} \frac{{\rm d} x}{x}~F_{\gamma/e}(x)
~F_{\gamma/e}(\frac{z^2}{x}), \label{cross}
\end{eqnarray}
where $F_{\gamma/e}$ denotes the energy spectrum of the
back-scattered photon for unpolarized initial electron and laser
photon beams given by
\begin{eqnarray}
F_{\gamma/e}(x)&=&\frac{1}{D(\xi)} \left ( 1-x+\frac{1}{1-x}-\frac{4
x}{\xi (1-x)}+ \frac{4 x^2}{\xi^2 (1-x)^2} \right ).
\end{eqnarray}
The definitions of parameters $\xi$, $D(\xi)$ and $x_{max}$ can be
found in \cite{distribution}. In our numerical calculation, we
choose $\xi=4. 8$, $D(\xi)=1. 83$ and $x_{max}=0. 83$.

\subsection{Amplitudes for $\gamma\gamma \to \bar \ell_i \ell_j$}
\begin{figure}
\begin{center}
\includegraphics [scale=0.7] {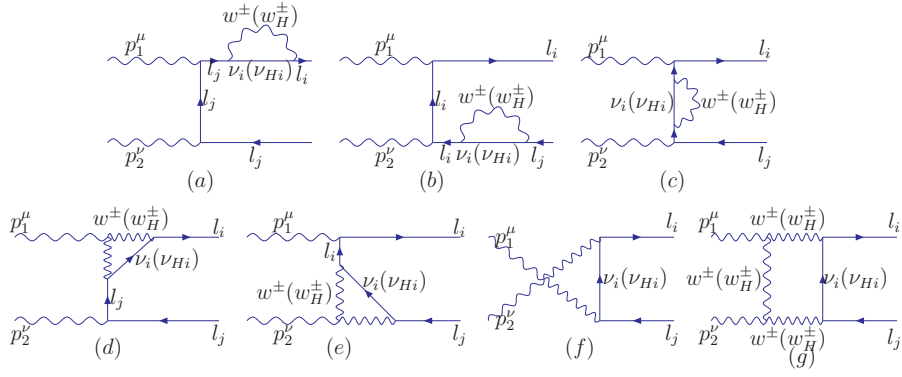}
\caption{Feynman diagrams for the production
$\gamma\gamma\to \bar \ell_i \ell_j$ in the LRTH model mediated
 by the heavy and light gauge bosons $W_{L,R}^\pm$. Those with
the two photon lines crossed are not shown.  } \label{fig1}
\end{center}
\end{figure}

\begin{figure}
\begin{center}
\includegraphics [scale=1] {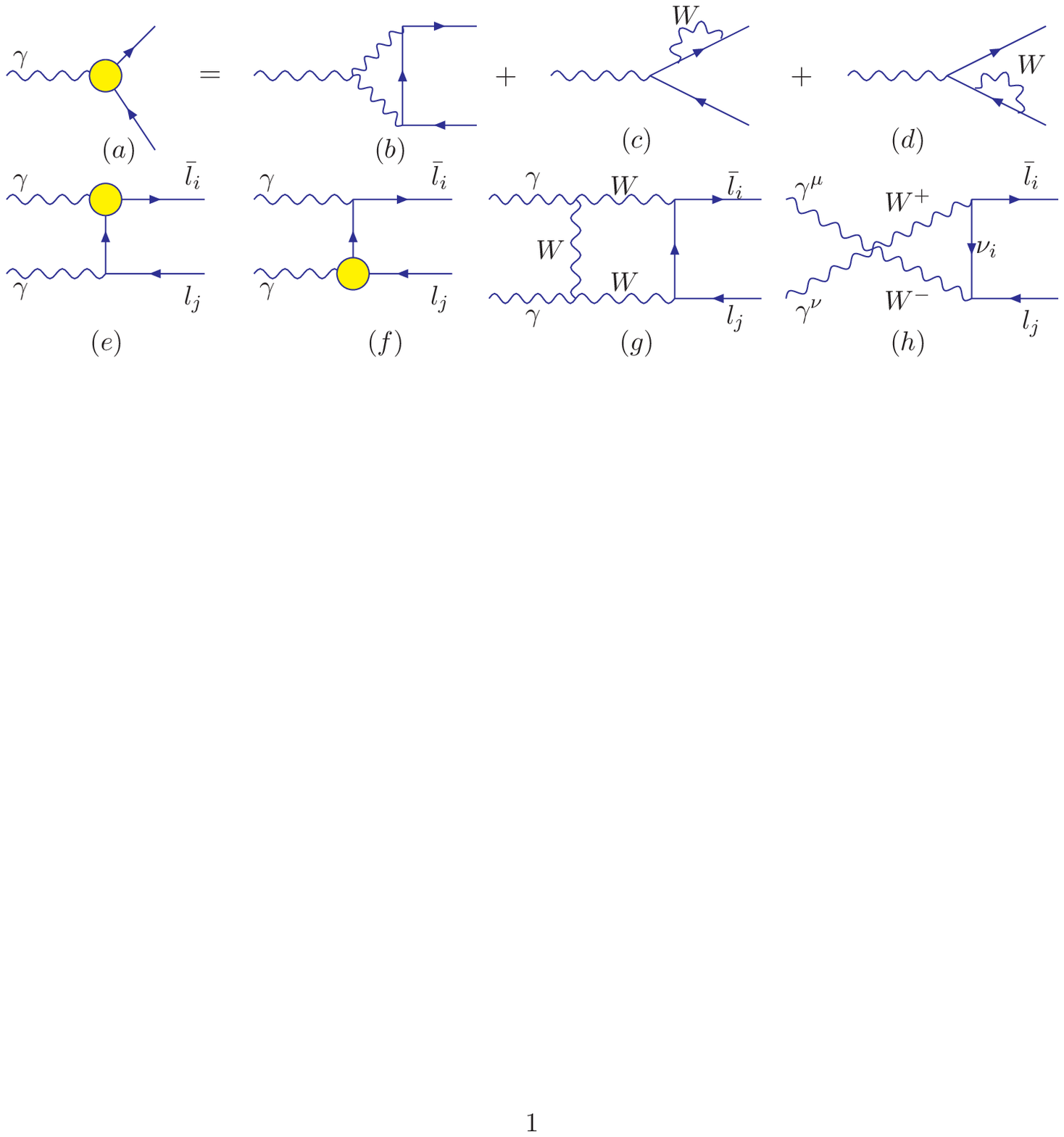}
\caption{Feynman diagrams for the production
$\gamma\gamma\to \bar \ell_i \ell_j$ in the LRTH model, with the triangle- and the self-energy-diagrams replaced by the tree level vertex (a), i.e., (b)(c)(d).  } \label{fig2}
\end{center}
\end{figure}

Via the coupling in Eq.(\ref{hlv}), the Feynman diagrams for the production $\gamma \gamma
\to \bar \ell_i \ell_j$ mediated by the charged gauge bosons are shown in Fig. \ref{fig1}. The
contributions from the charged scalars have the similar structure as that from the gauge boson.
That is, if the boson lines change into scalar lines in Fig. \ref{fig1} and Fig. \ref{fig2}, they will
become to the Feynman diagrams contributed by the charged scalars, which have not shown explicitly.

It can also be seen that we have changed  Figs.1(a), 1(b), 1(c), 1(d) and 1(e) into Figs.2(e) and 2(f) via extracting a vertex shown as Figs.2(a), 2(b), 2(c) and 2(d)\cite{0702264}. To obtain this, we split the propagator in Fig.\ref{fig1}(c)
into two parts:
\begin{eqnarray}
M_c  \propto \frac{i} {\q_slash - m_i} i \Sigma(q)
\frac{i}{\q_slash - m_j} = \frac{i (\q_slash + m_i )}{m_j^2 -
m_i^2} \ i \Sigma(q) \ \frac{i}{\q_slash - m_j} +
 \frac{i}{\q_slash- m_i} \ i \Sigma (q) \  \frac{i (\q_slash + m_j)}{m_i^2 -
 m_j^2}.   \label{technique}
\end{eqnarray}
In the right-handed terms of Eq. (\ref{technique}), the first term together with Fig.1 (a, d),
and the second term together with Fig.1(b, e) can be collected into a vertex, irrespectively.
Then the momentum dependent $\bar \ell_i \ell_j \gamma$ vertex, after this arrangement, can be
defined as,
\begin{eqnarray}
\Gamma^{'\bar \ell_i \ell_j \gamma}_{\mu} (p_i, p_j) &= & \Gamma_\mu^{\bar \ell_i \ell_j \gamma}
(p_i,p_j) + i \Sigma (p_i)   \frac{i (\p_slash_i + m_j)}{m_i^2 -
m_j^2} \Gamma_\mu^{\bar{l}l' \gamma} + \Gamma_\mu^{\bar{l}l'\gamma} \frac{i
(\p_slash_j + m_i )}{m_j^2 - m_i^2} \ i \Sigma(p_j), \label{eff}
\end{eqnarray}
where $\Gamma_\mu^{\bar \ell_i \ell_j\gamma} $ is the penguin diagram
contribution to the total $\bar \ell_i \ell_j \gamma$ vertex,
then the calculation of Fig.\ref{fig1} (a-e) is
equivalent to the calculation of the "tree" level process
depicted in Fig.\ref{fig2} (a) and (b), which obviously has a
simpler structure.

As for the calculation of the $\bar \ell_i \ell_j \gamma$ vertex, we can firstly give the results from the Lorentz structure,
To discuss the contribution of the self energy diagrams, we take Fig.\ref{fig2}(c) as an example, and
 the amplitude can be written as,
 \beq
 {\cal M}_c \sim \gamma^\rho \frac{1}{\p_slash-\k_slash-m_{\nu_H}}\gamma_\rho \frac{1}{\p_slash} \gamma^\mu
 \cdot \bar \ell_i \ell_j \epsilon_\mu.
 \eeq
 The electromagnetic gauge invariance $\partial_\mu {\cal M}=0$ has required this term vanishing. So does Fig.\ref{fig2}(d).

So there is only the Fig.\ref{fig2}(b) left. When we sum over all the diagrams corresponding to the three intermediate mass
eigenstate, note that,
\bea
\sum\limits_i\{\frac{ U^*_{ei}U_{\mu i}}{(p+k)^2-m^2_{\nu_H}}\}
&=& \nonumber
\sum\limits_i U^*_{ei}U_{\mu i}\{ \frac{1}{(p+k)^2}+\frac{m_i^2}{[(p+k)^2]^2} +...\} \\
&=&\sum\limits_i\frac{  U^*_{ei}U_{\mu i}m^2_{\nu_H}} {[(p+k)^2]^2 }+...,
\eea
the leading term vanishes via the GIM mechanism, $\sum\limits_i U^*_{ei}U_{\mu i}=0.$ The Second term, with more
powers of $k$ in the denominator, has already cleared away the UV divergence.

The penguin contributions from the heavy gauge bosons
and the charged scalars in unitary gauge ($\xi\to \infty $), which are calculated by hands,
via Feynman parameterization and Wick rotation, can be written
as\cite{l-f-li}
\bea
{\cal M}_{W_H}  &=&\frac{ce^3}{(\sqrt{2}s_W)^2}\frac{m_i}{64\pi m_{W_H}^4}\bar u_i(p)(1-\gamma_5)
(2p\cdot \epsilon - m_i\gamma\cdot \epsilon) u_j(p-k) \\
{\cal M}_{H_\pm}&=&- 2e \frac{cm_i}{32\pi f^2 m_{H}^2}\bar u_i(p)(1-\gamma_5)(2p\cdot \epsilon - m_i\gamma\cdot
\epsilon) u_j(p-k) \label{am_h}
\eea
where $c=\sum\limits_i U^*_{ei}U_{\mu i} m^2_{i\nu_H}$ and $m_{i\nu_H}$ is the ith generation heavy neutrino mass.
$p,~k$ is the momentum of production heavier lepton
and the photon of the vertex, respectively,
and $m_i $ is the heavier lepton mass.

As for the box diagram Fig.\ref{fig2}(g) and the bosonic quadruple interaction in Fig.\ref{fig2}(h),
we have use the calculating tool of LoopTools\cite{looptools}.

\section{Numerical results}
In our calculations, we neglect terms proportional to $v^2/f^2$ in
the new gauge boson masses and also in the relevant Feynman rules.
We take the SM parameters as \cite{pdg-2016}:
\begin{eqnarray}
&& m_{e}=0. 0051\ {\rm GeV}, \quad \quad  m_{\mu}= 0. 106\ {\rm GeV},
\quad \quad  m_\tau= 1. 777\ {\rm GeV}, \nonumber \\ &&  m_Z =91. 2\
{\rm GeV}, \quad \quad \quad s^{2}_{W}= 0. 231, \quad \quad
\alpha_e=1/128. 8.  \nonumber
\end{eqnarray}
The internal charged lepton masses, $m_e,~m_\mu,~m_\tau$, however, will
be neglected since they are much lighter than the gauge bosons, the charged scalars,
or the right-handed neutrinos. 

  When the gauge boson is mediated in the loop, just as shown in Fig. \ref{fig1} and Fig. \ref{fig2},
   the relevant parameters are the masses of the gauge bosons $m_W,~m_{W_H}$
and the heavy neutrino $m_{\nu_H}$. 
On the other side, the heavy charged bosons may also contribute large to the lepton flavor changing processes,
which can be realized by replacing the heavy gauge bosons with the charged scalars $\phi^\pm$ in Fig.\ref{fig1} and Fig. \ref{fig2}.

In the Higgs mediated process, in addition to the masses of the charged scalars $m_\phi $
and the heavy neutrino $m_{\nu_H}$, the breaking scales $f$, $\hat f$ are also dependent parameters.
The light neutrino masses and the charged leptons mixings to the light neutrinos $c_i$ ($\phi^-\bar l \nu_{L,R}$)
are quite small, so we here neglect the contributions mediated by the light neutrinos.
We will focus on the heavy neutrinos, which coupling to charged leptons via the charged scalars is
 proportional to the heavy neutrino mass, i.e, $\sim c_H\frac{\hat f^2}{\Lambda f}$.

For the masses of the charged scalars and the heavy gauge bosons, we vary their ranges as:
$200 \leq m_\phi  \leq 1000$ GeV \cite{charged_higgs_mass} (sometimes, extending to $100$ GeV)
and $1000 \leq M_{W_H} \leq 5000 $ GeV \cite{w_h_constraints}.


Note that in the couplings of $\phi^+(W_H^+)\nu_H^k \bar \ell$ there exist the mixing terms $V_H^{kl}$s, which
parameterize the interactions of the charged leptons with the heavy neutrinos, mediated by both $\phi^\pm$ and $W_H^+$,
and they can be chosen as the Maki-Nakagawa-Sakata (MNS) matrix $V_\MNS$, which
diagonalizes the neutrino mass matrix mass\cite{Maki:1962mu,07124019}:
\begin{equation} \label{vmns}
V_\MNS^{} =
\bmaT
c_{12}c_{13}                        & s_{12}c_{13}                  & s_{13}e^{-i\delta} \\
-s_{12}c_{23}-c_{12}s_{23}s_{13}e^{i\delta}  & c_{12}c_{23}-s_{12}s_{23}s_{13}e^{i\delta}  & s_{23}c_{13} \\
s_{12}s_{23}-c_{12}c_{23}s_{13}e^{i\delta}   & -c_{12}s_{23}-s_{12}c_{23}s_{13}e^{i\delta} & c_{23}c_{13}
\emaT
\,,
\end{equation}
where $s_{ij}\equiv\sin\theta_{ij}$ and $c_{ij}\equiv \cos\theta_{ij}$. $\delta$ is the CP-phase.

Three mixing angles $\theta_{12}$, $\theta_{13}$, $\theta_{23}$ can be chosen as free parameters since the
they are different from those of the SM.  The contribution of the CP-phase $\delta$, varying from $0\sim 2\pi$,
can be a free parameter.
But we take firstly the three mixing angles $\theta_{12}$, $\theta_{13}$, $\theta_{23}$
and the CP-phase $\delta$ as\cite{solar,atm,acc,daya-bay,delta}
\begin{eqnarray}
\sin^22\theta_{12}\simeq 0.86 \,,~~~~ \sin^22\theta_{23}\simeq 1 \,,~~~~
\sin^22\theta_{13}\simeq 0.089\,,~~~~\delta\simeq \pi,
\label{obs_para}
\end{eqnarray}
and in the final discussion we vary them as free parameters.


%



\def\figsubcap#1{\par\noindent\centering\footnotesize(#1)}
\begin{figure}[bht]%
\begin{center}
\hspace{-2.5cm}
 \parbox{5.05cm}{\epsfig{figure=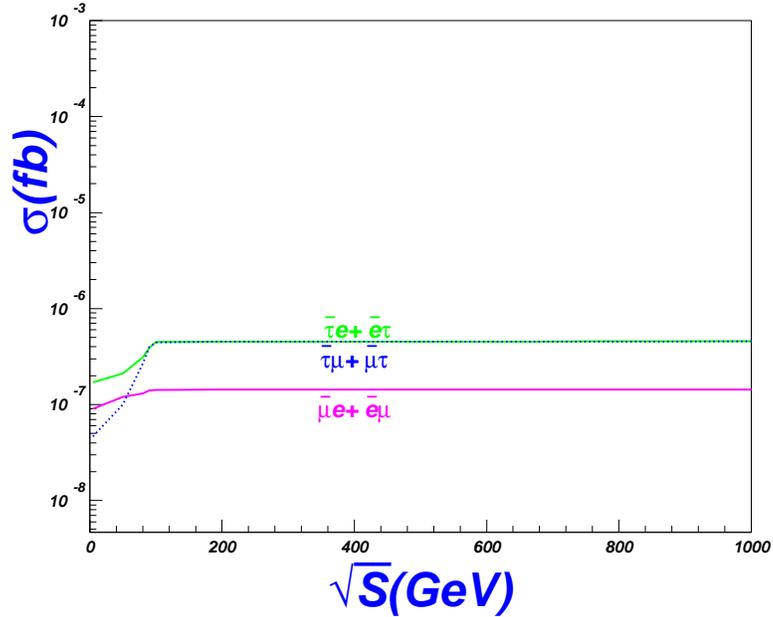,width=10.25cm} }
  \caption{ The cross sections of the process $\gamma\gamma$ $\to \bar \mu e + \bar e \mu$,
  $\to \bar \mu e + \bar e \mu$ and $\to \bar \mu e + \bar e \mu$ vary as the increasing centre-of-mass energy.
\label{fig3} }
\end{center}
\end{figure}

\subsection{The SM Background of the Flavor Changing Processes}
The SM backgrounds of the flavor changing production is quite small, since
these processes are prohibited in the tree level and suppressed largely in the one-loop level\cite{susy-r-con1}.
The main backgrounds of the
$\tau \bar e$ may be $\gamma\gamma \to \tau ^{+}\tau^{-} \to \tau \nu_{e}\bar{\nu}_{\tau}\bar e$,
$\gamma\gamma \to W^{+}W^{-}\to \tau \nu_\tau \nu_e \bar e $
and  $\gamma\gamma \to \tau \bar e \nu_\tau\nu_e$,  which
are suppressed to be $~9.7\times 10^{-4}$ fb, $~1.0\times 10^{-1}$ fb and
$~2.4\times 10^{-2}$ fb.
If $3.45 \times 10^2$ fb$^{-1}$ integrated luminosity of the photon collision \cite{tesla} is chosen,
the production rates of $\gamma \gamma \to \mu\bar e,~\tau\bar{e}, ~\tau \bar{\mu}$
should be larger than $10^{-2}$ fb to get
the $3 \sigma$ observing significance \cite{susy-r-con1,rvio-susy-rrll}.

\begin{figure}[bht]%
\begin{center}
\hspace{-2.5cm}
 \parbox{10.05cm}{\epsfig{figure=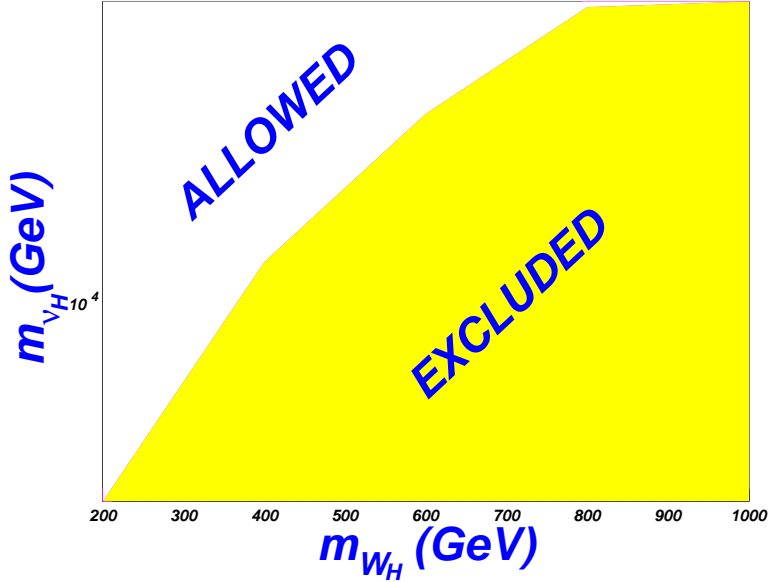,width=10.25cm} } 
  \caption{ The contour of the process $\gamma\gamma \to \bar \mu e + \bar e \mu$, 
           between  $m_{W_H}$ and  $m_{\nu_H} $. 
\label{fig4} }
\end{center}
\end{figure}

In the calculation, to avoid the collinear divergence, we require that the scattering angle cut $|\cos\theta_e|<0.9$
and the transverse momentum cut $p^e_{T}>20{\rm ~GeV}$, which are the same as the cuts in Ref. \cite{susy-r-con1}.
Therefore the requirement of the cross section $10^{-2}$ fb can be used to constraint the parameter
such as $f$, $m_\phi $, $m_{W_H}$ and $m_{\nu_H}$, etc and give the contours between them, just shown as Fig. (\ref{fig4}), Fig. (\ref{fig6}).

\def\figsubcap#1{\par\noindent\centering\footnotesize(#1)}

\subsection{The Contour of $m_{W_H}$ and $m_{\nu_H}$ in the $W_H$-mediated process}
Since the relation of parameters in $\gamma\gamma \to \bar \ell_i \ell_j$ mediated by the heavy $W_H$ is a little simple, we will begin from
this channel to discuss the dependence of the parameters. Of course, the process $\gamma\gamma \to \bar \ell_i \ell_j$
should receive the contribution from
both the heavy gauge bosons and the charged scalars, and we will discuss this later.

To find the influence of the center-of-mass energy, we plot in Fig.\ref{fig3} that the cross section changing with the increasing
$\sqrt{S}$, and the results are in our expectation. We can see that the production rates of the three channels are almost in the same order,
and the trend of every channel is almost flat, so in our following discussion, we will take  $\sqrt{S}=200$ GeV and
neglect the minor difference induced by it.

From Fig.\ref{fig3}, we also see that the three curves in our precision range are almost the same, at least in the same order, so we
in the followings will only consider one process, for example, the  $\bar \mu e + \bar e \mu$ production.

We will give the contour of $m_{W_H}$ and $m_{\nu_H}$ firstly in Fig. \ref{fig4},
in which, the $W_H$ is taken between $200$ and $1000$ GeV, but in actual case, we should have
a larger $m_{W_h}$, e.g., larger than $1000$ GeV, so we can conclude that if $10^{-2}$ fb limit is assumed,
the possibility for $m_{W_H}$ and $m_{\nu_H}$ to survive together is quite small.


\subsection{The Contributions from the $f$, $m_\phi $, $m_{\nu_H}$ and $m_{W_H}$}


The VEVs $f$ and $\hat f$ of the two Higgses $H$ and $\hat H$, respectively,
are taken as $ 500\leq f\leq 5000$ and $\hat f = 10f$ in this work \cite{0711.1238}.
The parameters mainly involved are the parameters $m_{W_H}$, $m_\phi $, $m_{\nu_H}$, the Higgs VEV $f$,
 and the mixing matrix $V_H$,  which will be emphatically discussed.

We show in Fig.\ref{fig5} the dependence of parameters $f$, the scalar mass $m_\phi $, the heavy neutrino mass $m_{\nu_H}$ and the
heavy chaged boson mass $m_{W_H}$. We also see from in Fig.\ref{fig5} that the dependence of $f$, $m_{\nu_H}$,  $m_\phi $ and $m_{W_H}$ is
large enough to be detectable in some parameter space, for the $10^{-2}$ limit, with the requirements:
$f< 1400 GeV, ~m_{\nu_H}> 6000 GeV$ and looser $m_\phi,~m_{W_H}$.

We notice that in Fig.\ref{fig5} (a)(b), the $f$ and the $m_{\nu_H}$ dependence, which have opposite influence on the production rates,
i.e., the cross section is increasing with a increasing $m_{\nu_H}$, but a decreasing $f$,
which can be understandable since from Eq.(\ref{hlv}),
we can see that the coupling of $\phi \bar l\nu_H $ proportional to $m_{\nu_H}$, while inverse proportional to $f$.

\def\figsubcap#1{\par\noindent\centering\footnotesize(#1)}
\begin{figure}[bht]
\begin{center}
\hspace{-2.5cm}
 \parbox{7.05cm}{\epsfig{figure=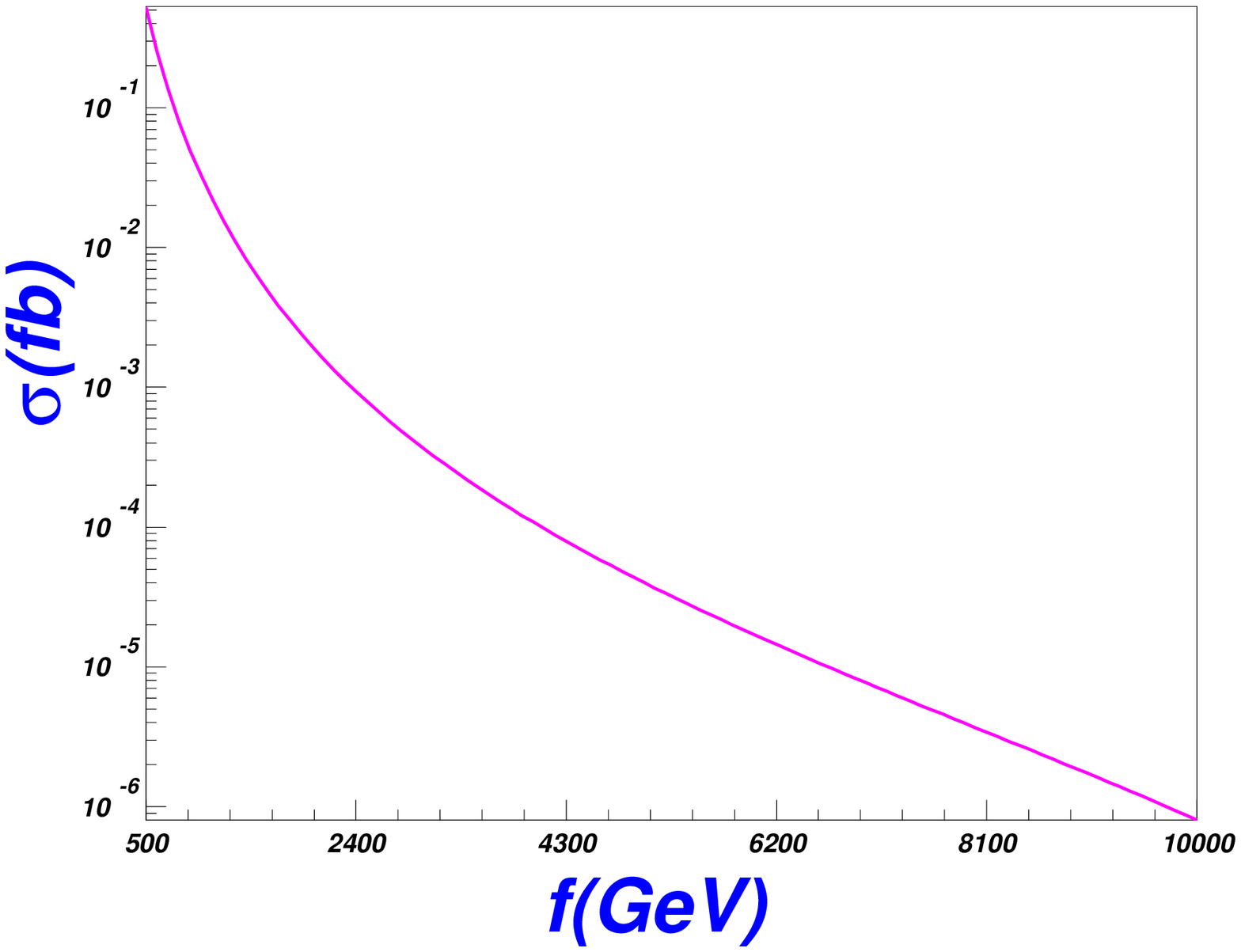,width=7.25cm} \vspace{-1.2cm}\figsubcap{a} }
 \parbox{7.05cm}{\epsfig{figure=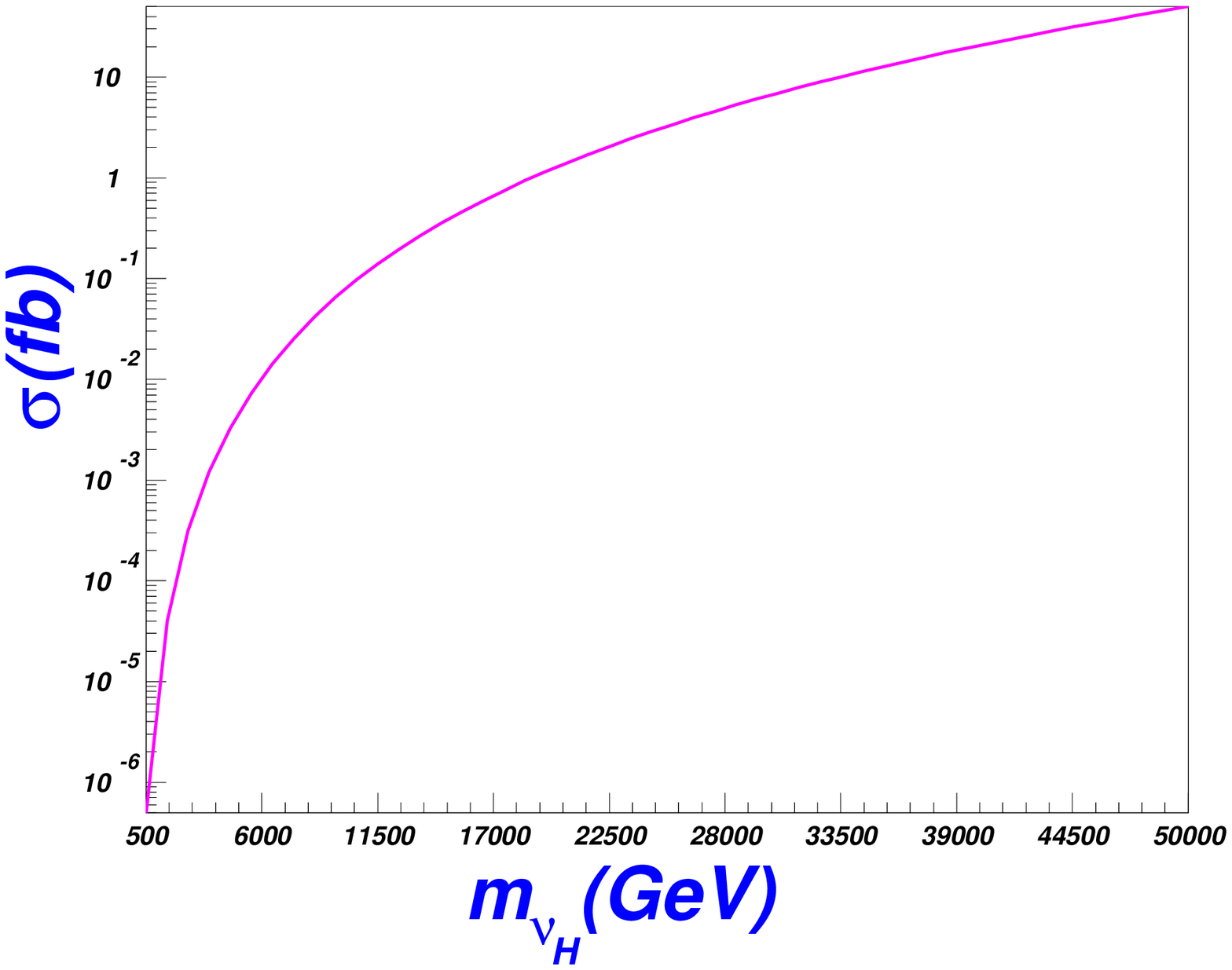,width=7.25cm} \vspace{-1.cm}\figsubcap{b} } \vspace{0.2cm} \\ \hspace{-2.5cm}
 \parbox{7.05cm}{\epsfig{figure=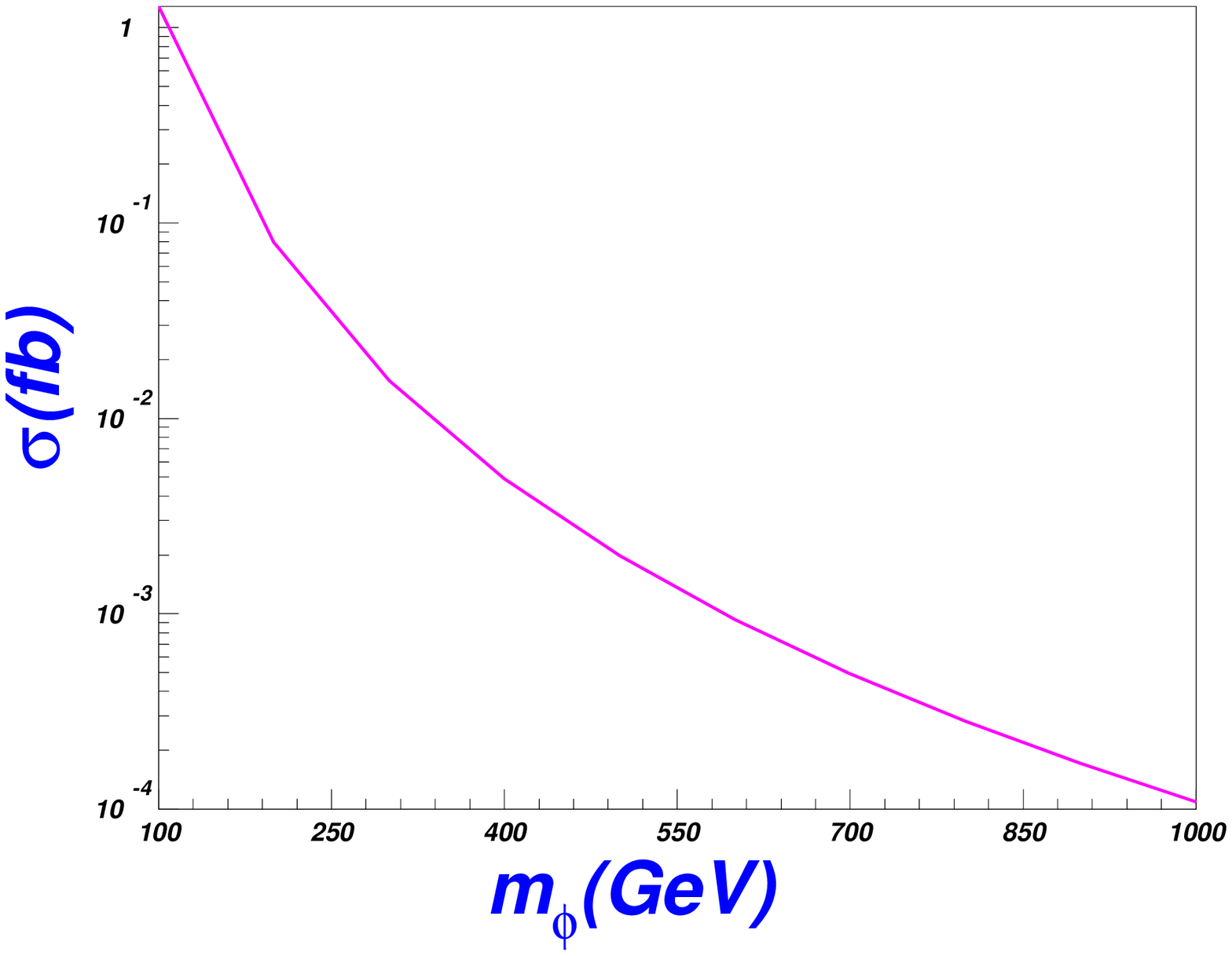,width=7.25cm} \vspace{-1.2cm}\figsubcap{c} }
 \parbox{7.05cm}{\epsfig{figure=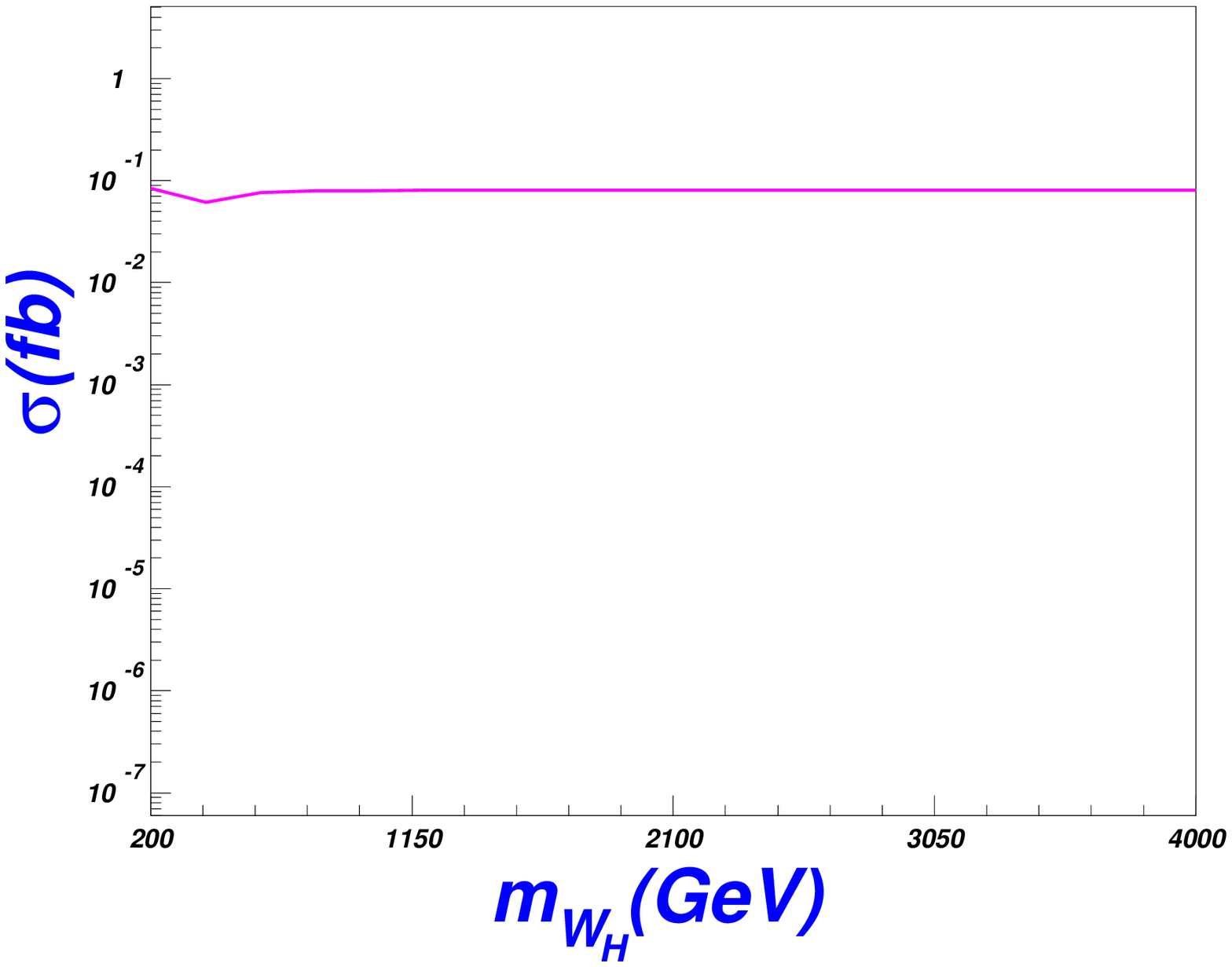,width=7.25cm} \vspace{-1.cm}\figsubcap{d} }
   \caption{ The cross section $\sigma$ of the processes $\gamma\gamma
         \to  \bar \tau \mu $ as a function of the breaking $f$, the scalar mass $m_\phi $, the heavy neutrino mass $m_{\nu_H}$ and the
         heavy charged boson mass $m_{W_H}$, respectively.
\label{fig5} }
\end{center}
\end{figure}

\def\figsubcap#1{\par\noindent\centering\footnotesize(#1)}
\begin{figure}[bht]
\begin{center}
\hspace{-2.5cm}
 \parbox{7.05cm}{\epsfig{figure=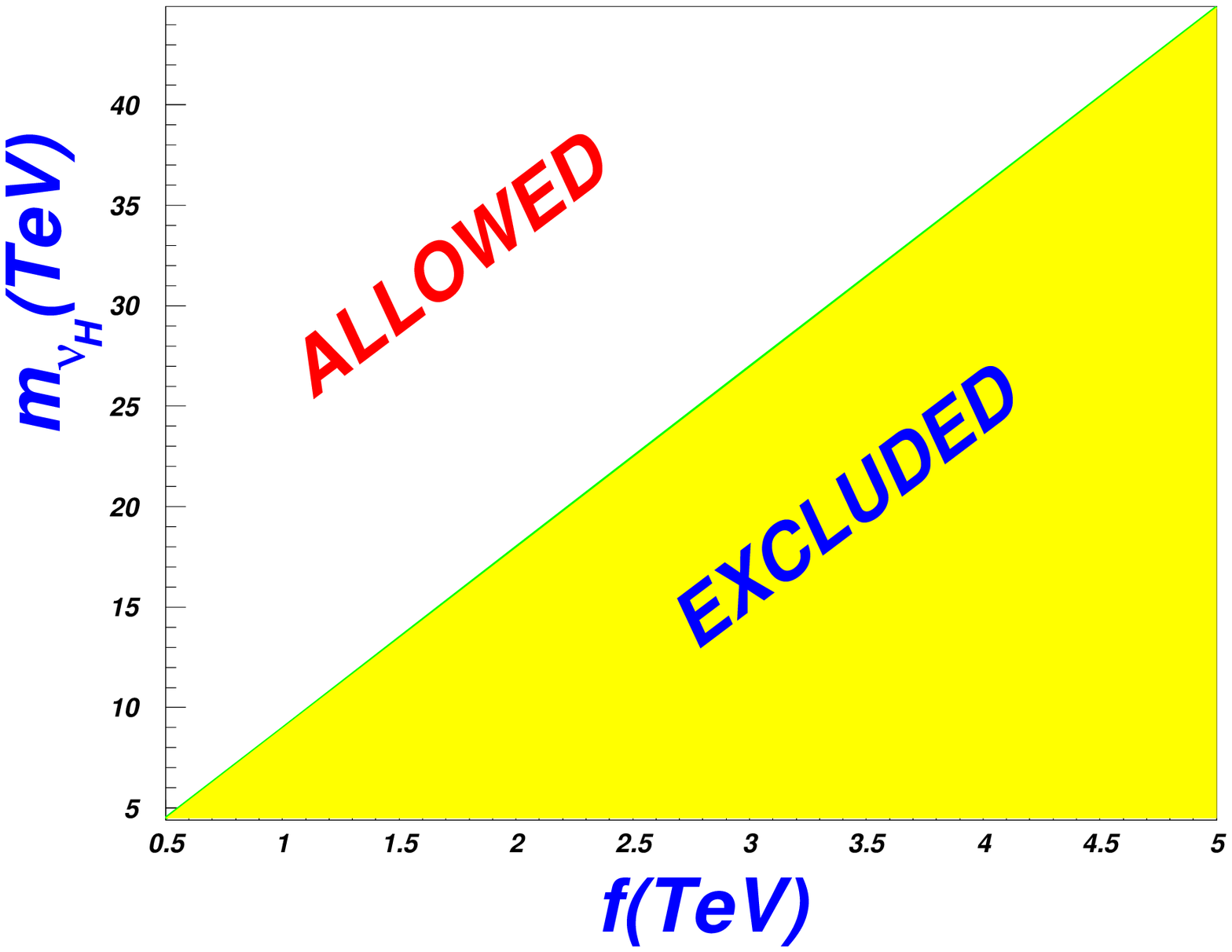,width=7.25cm} \vspace{-1.2cm}\figsubcap{a} }\hspace{1cm}
 \parbox{7.05cm}{\epsfig{figure=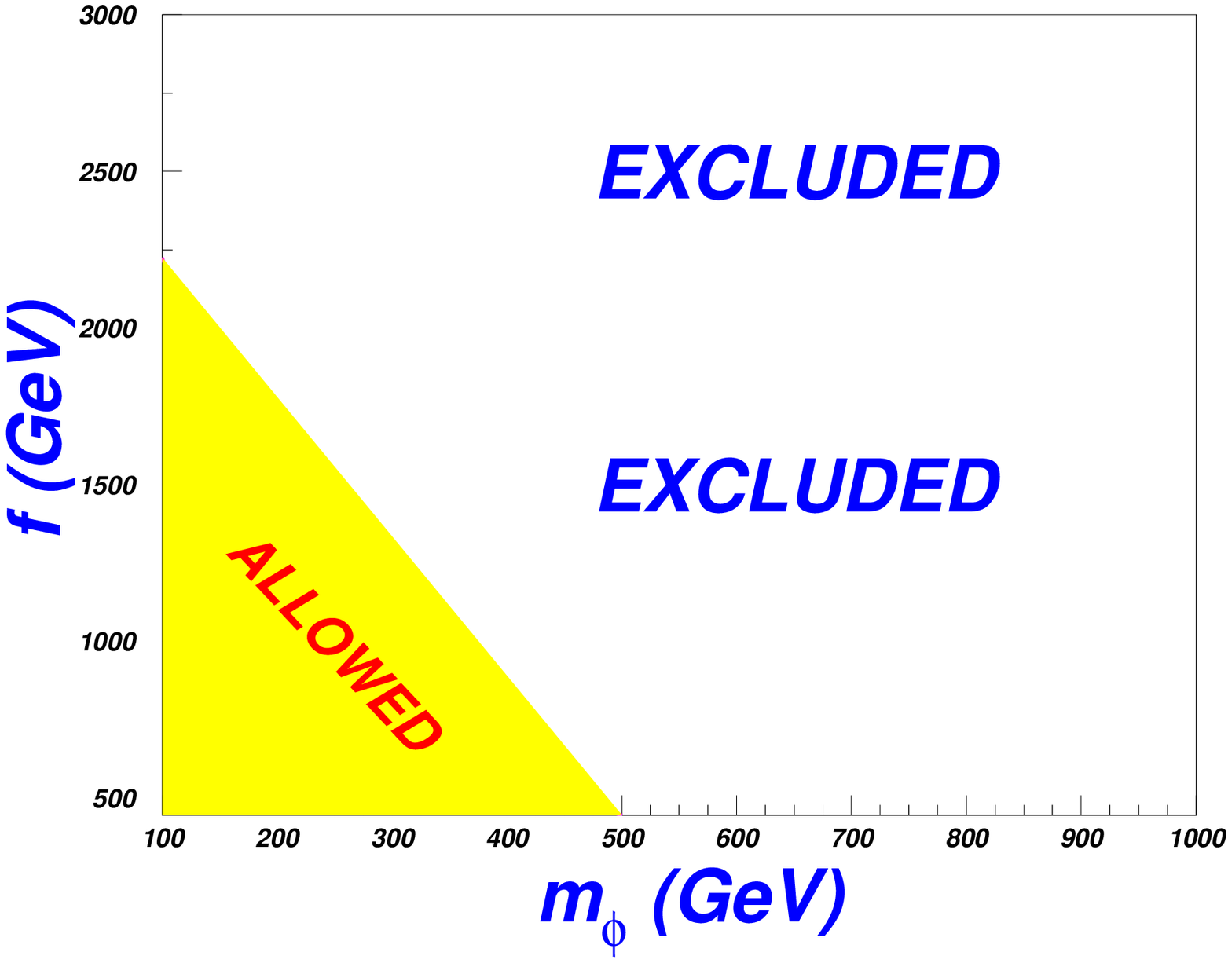,width=7.25cm} \vspace{-1.2cm}\figsubcap{b} }\hspace{-3cm}
 \parbox{9.05cm}{\epsfig{figure=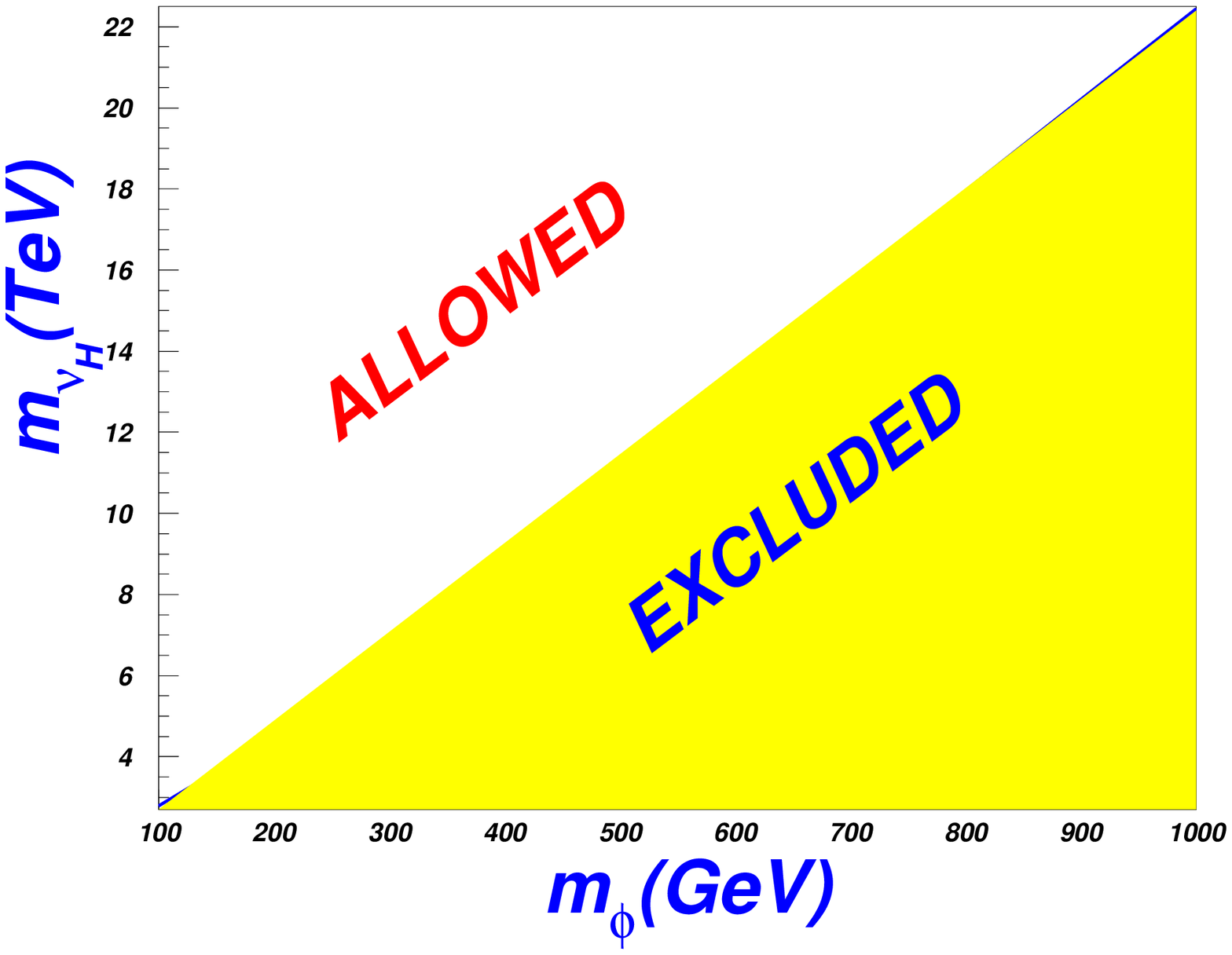,width=9.25cm} \vspace{-1.2cm}\figsubcap{c} }
    \caption{ The contour of $f$ and $m_{\nu_H}$(a), of $m_\phi $ and $f$(b), and of $m_\phi $ and $m_{\nu_H}$ (c) \label{fig6} }
\end{center}
\end{figure}

\def\figsubcap#1{\par\noindent\centering\footnotesize(#1)}
\begin{figure}[bht]
\begin{center}
\hspace{-2.5cm}
 \parbox{7.05cm}{\epsfig{figure=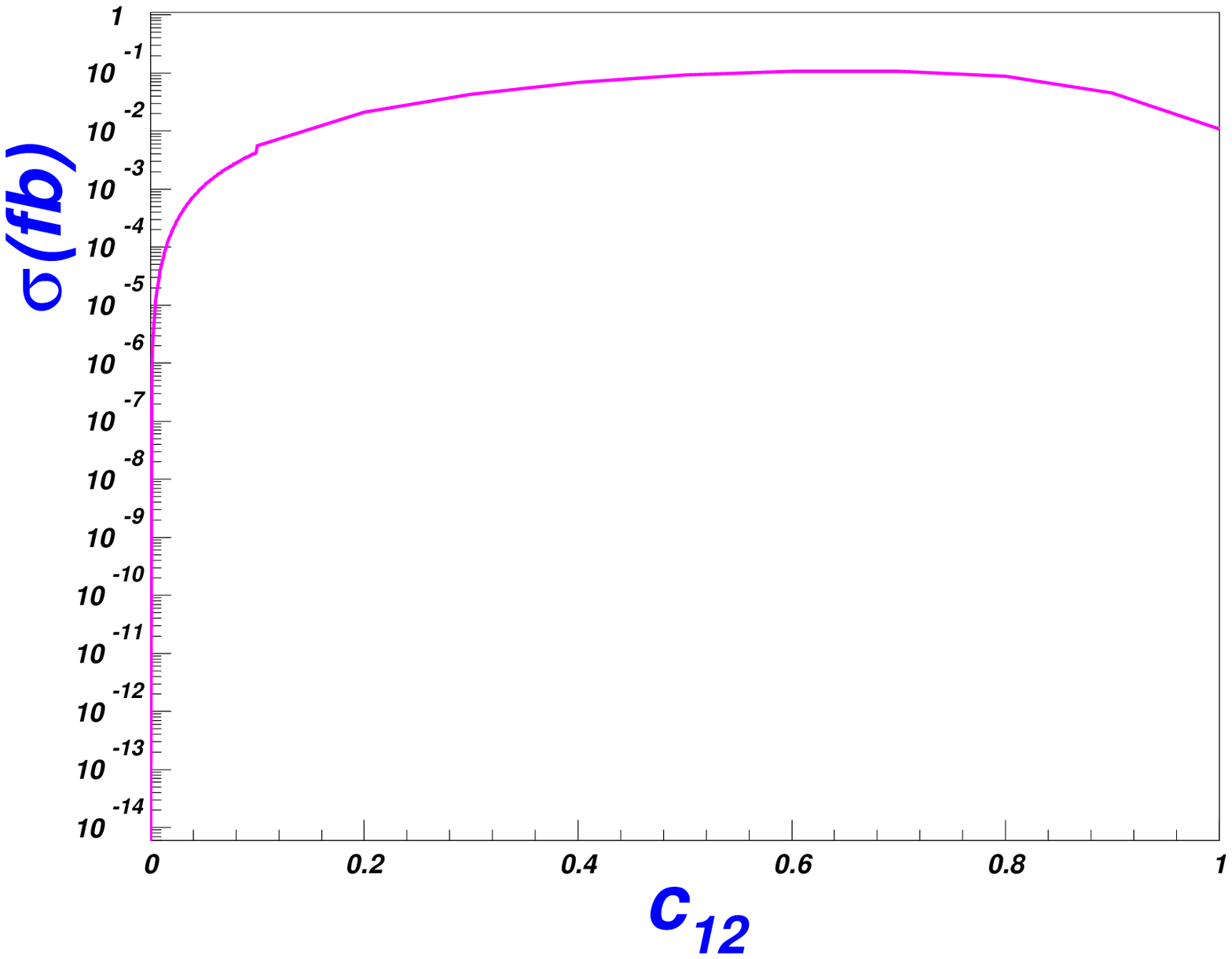,width=7.25cm} \vspace{-1.2cm}\figsubcap{a} }
 \parbox{7.05cm}{\epsfig{figure=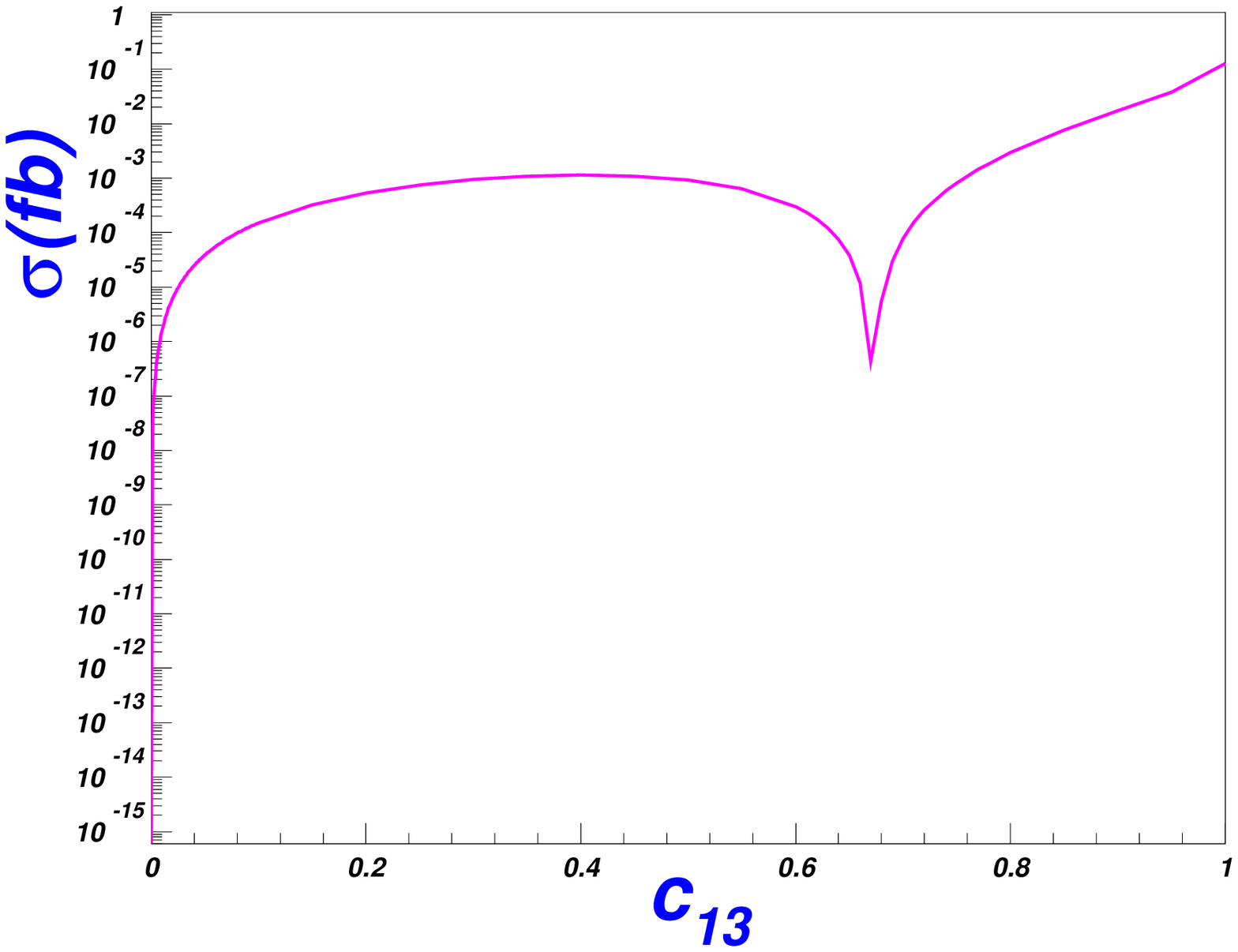,width=7.25cm} \vspace{-1.2cm}\figsubcap{b} } \\
 \hspace{-2.5cm}
 \parbox{7.05cm}{\epsfig{figure=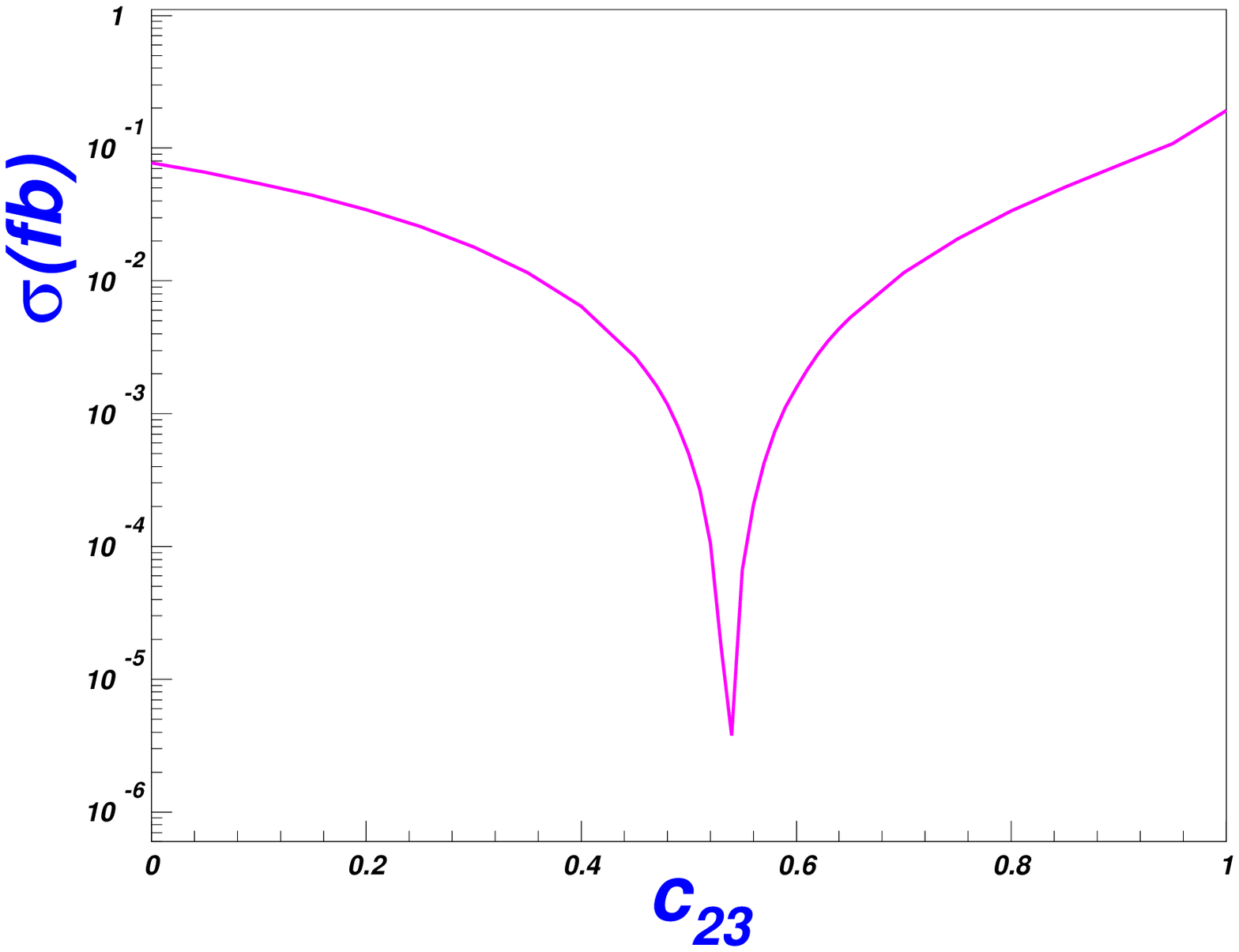,width=7.25cm} \vspace{-1.2cm}\figsubcap{c} }
 \parbox{7.05cm}{\epsfig{figure=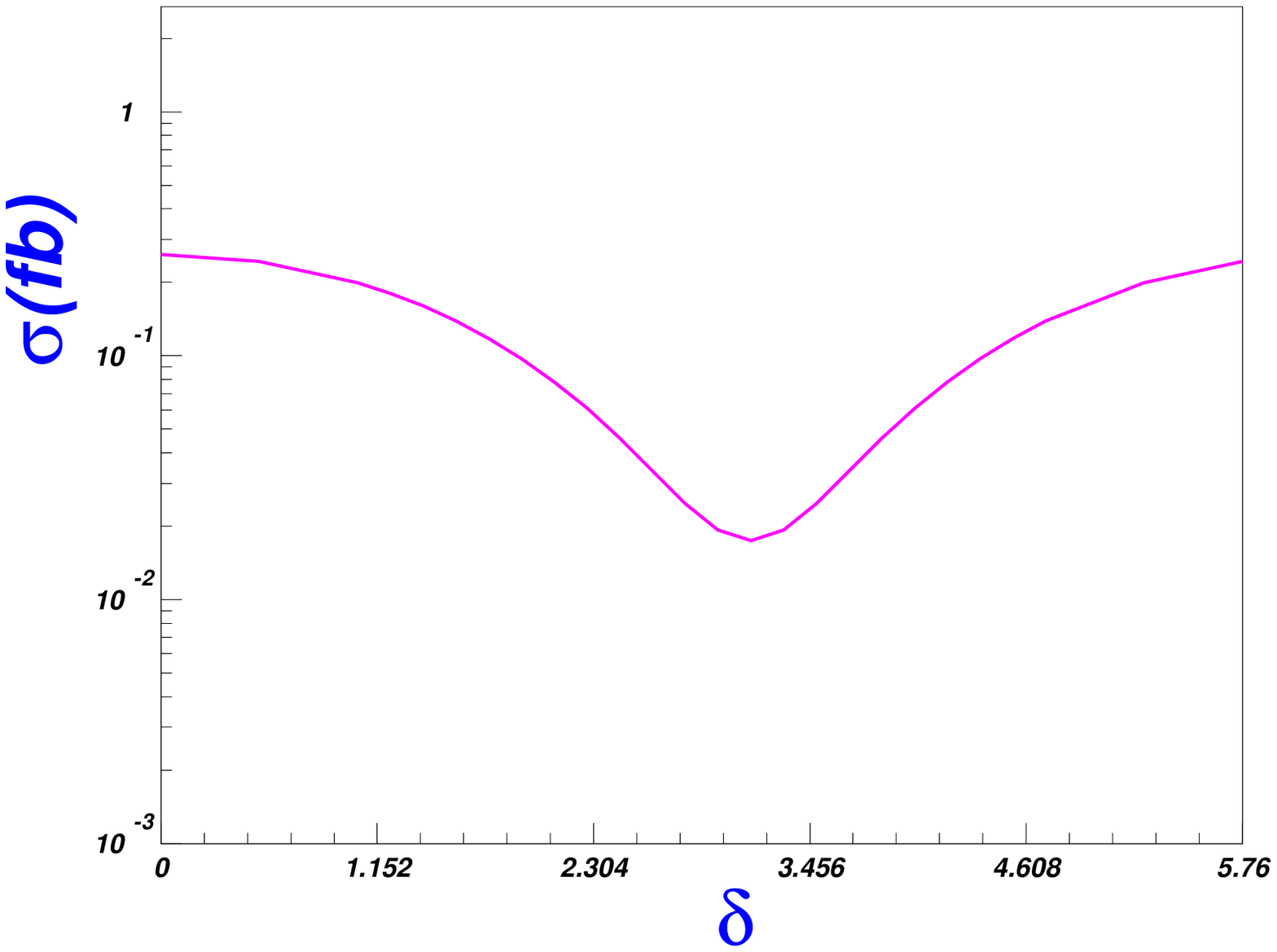,width=7.25cm} \vspace{-1.2cm}\figsubcap{d} }
    \caption{The production rates as functions of $c_{12}$, $c_{13}$, $c_{23}$, and $\delta$.  \label{fig7} }
\end{center}
\end{figure}
The production rates with the $m_\phi $ in  Fig.\ref{fig5}(c), are large, and the total range of
the vertical axis is not too wide: 0.009 -0.018 fb,
which provide the possibility to measure the scalar mass.

From Fig.\ref{fig5} (d), we can see that the $m_{W_H}$ dependence seems quite large, the cross sections, however,
the contributions of $m_{W_H}$ and $f$ are not related with each other, since the couplings of $W_H\bar l\nu_H $
in Eq.(\ref{wlv}) do not comprise the breaking parameter $f$, so in Fig. \ref{fig5}(d) the curve of the cross section
on $m_{W_H}$ is flat especially when $m_{W_H}$ becomes large, which is because in the total production, the scalar
contribution dominates so that the change of the heavy gauge boson mass can not affect the production order.

Since in Fig.\ref{fig5} the dependence of $f$, the scalar mass $m_{H}$ and the heavy neutrino mass $m_{\nu_H}$
is large, Fig.\ref{fig6} will show the contour of $m_{\nu_H}$ vs. $f$ (a), $f$ vs. $m_{H}$(b)  and $m_{\nu_H}$ vs. $m_{H}$(c).

In Fig. \ref{fig6}(a)(c) we can see that the two contours have similar trend with the changing $m_{\nu_H}$.
 With increasing $m_{\nu_H}$, the cross section will increase too, so a large $m_{\nu_H}$
 is favored. We also see in Fig. \ref{fig6} that in our grossly discussion, if $f > 1000$ GeV,
 for the rates to arrive at the detectable production rates, $m_{\nu_H}$ must
  be larger than $8190$ GeV, while the scalar mass should be smaller than $300$ GeV.

In Fig. \ref{fig6}(b) we see the contour between $m_\phi$ and $f$, and the surviving space is quite small, which is understandable
since the largest contribution comes from the mass of the heavy neutrino, and we take $m_{\nu_H}=1000$ GeV in Fig.6 (b), which is
not enough to obtain a big production rate, so to arrive at the required cross sections, $f$ or $m_\phi$ should not too large, which
limit them in a small possible space.

From Fig. \ref{fig6}(a)(b)(c), we see that the right-handed neutrino mass contributes largest to the cross section, so this process may
serve as a severe constrain to the mass of the heavy neutrino.

Although we have discussed the dependences on $m_{W_H}$, $f,~ m_\phi , ~m_{\nu_H}$,
(see Figs. 4, 3, 5, and 6), we have not considered changing generation mixings, since we have
fixed them as the lepton mixing parameters [as in Eq. (\ref{obs_para})].
 In fig.7 we free them and plot the dependence of these mixing parameters.
 We find that the cross sections vary large in some ranges but in total are gradual, especially the curve of Fig.\ref{fig7}(d).

 In Fig.\ref{fig7}(a)(b), there are sharp points when $c_{12}=0$ or $c_{13}=0$, and we can find the reason in the expression
 of the mixing matrix in Eq.(\ref{vmns}), in which, the element $V_{12}$ and $V_{13}$ are proportional to $s_{12}$ and $s_{13}$, respectively.
 $c_{12}=0$ or $c_{13}=0$, $s_{12}=1$ or $s_{13}=1$ will contribute quite large.

\section{Conclusion}

Charged scalar- and gauge boson- mediated lepton flavor changing productions
of $\bar \ell_i \ell_j$ ($i\neq j$)  via $\gamma \gamma$ collision at
the ILC have been performed. We find that in a certain parameter space,
the production rates of $\gamma \gamma \to \bar \ell_i \ell_j$ ($i\neq j$) may
arrive at $10^{-2}$ fb, which means that we may have serval events each year for
the designed luminosity of about $345$ fb$^{-1}$/year at the ILC.
Due to the negligible observation of such $\bar \ell_i \ell_j$ events
in the SM, it would be a detection to the left-right twin Higgs models in
the lepton sector.

And more important, if we cannot detect the process, this may constrain the
parameters strictly. For example, if the process is undetectable, we can give a upper limit of the
Higgs breaking scale $f$. We can see from Fig.\ref{fig5}(a), to arrive at the cross section $10^{-2}$ fb, $f$
should be less than $1.4$ TeV in the set parameter space.

Moreover, since the LFV couplings are closely related to the heavy neutrino masses, we may obtain
interesting information for the heavy neutrino masses if we could see any signature of the LFV processes.
In Fig.\ref{fig5}(b), to arrive at the cross section $10^{-2}$ fb, the heavy neutrino mass $m_{\nu_H}$
should be larger than $6$ TeV in the given parameter space.

Therefore, these LFV processes may serve as a sensitive probe and
 a strict constraint of this kind new physics models.


\section{Acknowledgments}  
This work was supported by Excellent Youth Foundation of Zhengzhou University under grands No. 1421317053
and 1421317054, and the Natural Science Foundation of China under grant No. 11605110.

\hspace{1mm}

\end{document}